\documentclass[pre,reprint]{revtex4-1}
\usepackage{amsmath}
\usepackage{graphicx}				
\usepackage{amssymb}

\newcommand*{\pd}[2]{\frac{\partial #1}{\partial #2}}
\newcommand*{\pdel}[2]{\frac{\delta #1}{\delta #2}}
\newcommand*{\Tr}[1]{\operatorname{Tr}\left[{#1}\right]}
\newcommand*{\avg}[1]{\left\langle{#1}\right\rangle}
\newcommand*{\norm}[1]{\left | #1 \right|}

\newcommand*{\Z}{\mathcal Z_\text{kin}}
\newcommand{\tens}[2]{\left(#1 (\cdot) #2\right)}
\newcommand{\pure}[1]{|#1\rangle\langle #1|}
\newcommand{\upM}[4]{\ensuremath{\left[ \begin{array}{c|c}
 {#1} & {#2} \\
 \hline
 {#3} & {#4}
 \end{array} \right]}}

\begin{document}

\title{ Thermodynamics of Maximum Transition Entropy for Quantum Assemblies}
\author{ David M. Rogers, University of South Florida}

\begin{abstract}
  This work presents a general unifying theoretical framework for quantum non-equilibrium
systems.
It is based on a re-statement of the dynamical problem
as one of inferring the distribution of collision events that move a system
toward thermal equilibrium from an arbitrary starting distribution.
Using a form based on maximum entropy for this transition distribution
leads to a statistical description of open quantum systems with strong parallels
to the conventional, maximum-entropy, equilibrium thermostatics.
A precise form of the second law of thermodynamics can be stated for
this dynamics at every time-point in a trajectory.
Numerical results are presented for low-dimensional systems
interacting with cavity fields.  The dynamics and stationary state are compared
to a reference model of a weakly coupled oscillator plus cavity supersystem thermostatted by
periodic partial measurements.
Despite the absence of an explicit cavity in the present model of open quantum
dynamics, both the relaxation rates and stationary state properties closely match
the reference.
Additionally, the time-course of energy exchange and entropy increase is given
throughout an entire measurement process for a single spin system.
The results show the process to be capable of initially absorbing heat when starting
from a superposition state, but not from an isotropic distribution.
Based on these results, it is argued that logical inference in the presence of
environmental noise is sufficient to resolve the paradox of wavefunction collapse.
\end{abstract}



\maketitle

\section{ Introduction}

  Few topics draw more interest and debate than the relation between
quantum mechanics, the measurement problem, and information theory.
The problem of sewing these together into a coherent description of
a causal or probabilistic universe lies at the foundations of physical law.

  These long-standing topics are receiving renewed attention
due to the increasing precision of experiments on trapped
quantum systems\cite{dleib97,nherm00,qturc00,croos04,tkino06,dchan14} and
the corresponding ability to probe the nature of decoherence processes.\cite{qturc00a,lhack04,ghube08}
Parallel developments in nonequilibrium statistical mechanics of open systems
are building an increasingly powerful toolset for predicting
their often non-intuitive behavior.\cite{wzure02,mcamp11,tsaga13}
Important studied applications include spectroscopy\cite{nkhan03,ytani06},
quantum computing,\cite{ychen04}
reaction rate theory,\cite{wmill86}
mixed quantum-classical molecular dynamics\cite{rwyat01},
and superconducting nanocircuits.\cite{ndidi12}


  Recently, a general unifying theoretical framework for non-equilibrium thermodynamics was proposed
for classical systems.\cite{droge11a}
Such a framework has been a long sought goal for the nonequilibrium community.\cite{jlebo93,druel03}
It takes the form of a maximum entropy distribution over each transition step between
the current and future positions and momenta,
\begin{align}
P\left(x',p',t+dt\Big |x,p,t\right) = \Z^{-1} \delta((x'-x) - \tfrac{p'+p}{2M} \Delta t) \notag \\
 e^{-\pdel{S}{x(t)}^2 dt/2\sigma^2 + dt \dot x\pdel{S}{x(t)} \beta dt/ 2} \label{e:class}
.
\end{align}
The two constraints in the exponent are the average deviation of the square of the
action functional, and the average change in total energy.
Their Lagrange multipliers are $\sigma^{-2}/2$, one fourth the inverse diffusion
constant, and $\beta/2$, half the inverse thermal energy.
When the Hamiltonian is in the form, $H = p^2/2m + V(x)$ ($\pdel{S}{x(t)} = -dp/dt - \pd{V}{x}$), Eq.~\ref{e:class} delivers the usual Langevin picture of molecular kinetics,
where momentum updates are given by $dp = (F - \tfrac{\beta\sigma^2}{2}\dot x)dt + \sigma d\mathcal W$.
The {\em kinetic} partition function, $\mathcal Z_\text{kin}$, is a
moment generating function that contains all the information about force/flux relationships.
It makes an appearance in both the nonlinear fluctuation-dissipation
and entropy production relations.\cite{rstra92,droge11a}
It also provides a physical interpretation of the Lagrange multipliers (collision rate,
temperature, etc.) as constant forces responsible for bringing
about equilibrium through noisy energy exchange with external systems.
Equilibrium properties like temperature, pressure, or chemical potential are only defined
for the central system through the action of an external thermostat able to exchange
energy, volume, or molecules -- a consistent interpretation of temperature
out of equilibrium.\cite{jcasa03,droge12}

  Importantly, this work established that dynamic processes can serve as the
foundation for statistical mechanics.  In contrast to the fluctuation theorems, it
does not require Boltzmann-Gibbs equilibrium distributions (or even stationary states)
for initial and final densities.  Rather, the dynamics is determined directly from maximum
transition entropy.  Apart from being the stationary state of the dynamics,
the Boltzmann distribution plays no special role.
The log-ratio of forward and backward (inference) probabilities,
\begin{align}
\log &\frac{P\left(x',p',t+dt\Big |x,p,t,\beta\right) P(x,p,t)}{P\left(x,p,t\Big |x',p',t+dt,-\beta\right) P'(x',p',t+dt)}
\notag \\
 &= -\beta \Delta E - \log\frac{P'(x',p',t+dt)}{P(x,p,t)}
 ,\label{e:Sclas}
\end{align}
determines the `amount of irreversibility' for the entire process.  It consists of
changes in the information entropy of the system plus heat rejected to the environment.
The reverse process is determined by attempting to infer the starting distribution from
the final one -- a process subtly different from time reversal.
Most notably, since $\beta$ is a Lagrange multiplier favoring transitions with smaller
energy exchange, $\Delta E$, this changes sign during inference to
find starting states with larger energy.

  Based on the success of this work, it is reasonable to ask whether these results
can be extended to describe quantum dynamics.  The result would be a theory
that directly models the statistics of transitions between quantum states,
deriving thermodynamics as a consequence.
One major obstacle stands in the way of such an extension.  The quantum measurement problem
tangles up position and momenta, obscuring the classical picture of phase space.
This means that it is not possible to define a single path or trajectory which
a wavefunction undergoes to transition from one time point to the next.
This problem manifests in a variety of ways, including very
high frequency oscillations in the propagator for the Wigner distribution
and wavefunction collapse in matrix mechanics.

  This paper presents a novel statement of a transition probability distribution
over quantum states.  It retains much of the structure of the classical nonequilibrium
theory referred to above.  The distribution reduces to ordinary quantum
dynamics when a collision rate parameter is set to zero.
At nonzero rate parameter, the transition probability contains an exponential
bias toward trajectories representing exchange of energy with an external
thermostatic reservoir.  Additionally, a normalization constant for each step
of the dynamics acts as a moment generating function for the energy fluxes,
and a measure of irreversibility over a single timestep can be directly interpreted
in terms of heat and work.

  In what follows, we will prove the unique properties of this approach,
and analyze several of its consequences.  Section~\ref{s:deriv} provides
logical consistency arguments that eliminate alternative possible
transition probability assignments.  Section~\ref{s:ref} then introduces a novel quantum
Andersen thermostat acting on an explicit model of the environment in
an optical cavity experiment.
Section~\ref{s:math} goes over non-standard mathematical notation and properties
of transition superoperators used here.
Results in section~\ref{s:results} compare the simulation of our physically
motivated reference system (containing an explicit environment)
with the maximum transition entropy distribution (which contains no explicit environment).
Section~\ref{s:collapse} provides a detailed picture of energy and entropy
balance during a wavefunction collapse process which could occur
in a noisy interferometry experiment.  The conclusion discusses
interpretation questions and future extensions
and applications of the results.  The appendix contains a detailed
proof that the maximum transition entropy approach reduces
to the density-matrix master equation of Caldeira and Leggett in the
high-temperature limit.

\section{ Quantum Transition Processes}\label{s:deriv}

  Exact quantum dynamics follows unitary evolution,
\begin{equation*}
\psi(t) = U(t) \psi(0), \quad U(t) = e^{-\frac{i t}{\hbar}\hat H}
\end{equation*}
or, for the density matrix $\rho = \sum p_a |\psi_a \rangle \langle \psi_a |$,
\begin{equation*}
\rho(t) = U(t) \rho(0) U^\dagger(t)
.
\end{equation*}
The unitarity of $U^{-1} = U^\dagger$ means that both angles and distances between alternate
state vectors, $\psi_a$, remain unaltered during time evolution.  This also leads the eigenvalues
of $\rho$ to remain constant, so that entropy is a constant of the dynamics.

  A modification introduced by Jaynes\cite{ejayn57a} was to consider multiple
alternative, unitary time evolution operations,
\begin{equation}
\rho(t) = \sum_r P_r U_r(t) \rho(0) U_r^\dagger(t) \label{e:jaynes}
.
\end{equation}
The final state density matrix then becomes a mixture of outcomes due to
the unknown Hamiltonian that was applied to evolve the system in time.
Unfortunately, this mixing picture was left incomplete, since it could not
account for the temperature of the external system, and thus to
lead to the thermal distribution.

  For defining $U_r$, we can represent the effect of the external
system as a collection of random impulsive forces.
Each will cause the central system to undergo a
transition (in the position representation)
\begin{equation*}
\psi(x) \to e^{\frac{i}{\hbar}x\Delta p} \psi(x)
.
\end{equation*}
In the continuous limit, the index $r$ goes over into $\Delta p$.
These transitions are the short-time limit for any coupling
Hamiltonian linear in the coordinate, $x$.
Of course, other forms of environmental noise (e.g. periodic
or spatially localized functions) are possible.

Allowing for a quantum source of random impulsive forces, the most general transition operator will be the superposition,
\begin{equation*}
\rho' =
\int f(r) f^*(s) e^{-\frac{it}{\hbar} \hat H + \frac{i}{\hbar} r\hat x } \rho e^{\frac{it}{\hbar} \hat H - \frac{i}{\hbar} s\hat x} \; dr\; ds
.
\end{equation*}
Measuring the amount of momentum actually exchanged will collapse the
combined system onto a definite value of $r$, so that we need only consider noisy
transformations of the form (Eq.~\ref{e:jaynes}),
\begin{equation}
\rho' = \int |f(r)|^2 e^{-\frac{it}{\hbar} (\hat H - r\hat x/t)} \rho e^{\frac{it}{\hbar} (\hat H - r\hat x/t)} \; dr
.\label{e:noise}
\end{equation}

  Jaynes analyzed this propagator in detail for the case when $P_r$
(equivalently $|f(r)|^2$) does not depend on $\psi$ or $\rho$.\cite{ejayn57a}
We will term this case the ``unweighted'' distribution.
One central result is the subjective $H$ theorem,
\begin{equation}
S(t) \le S(t') \le S(t) -\sum_r P_r \log P_r
,\label{e:Hthm}
\end{equation}
For the increase in the von-Neumann entropy,\cite{vvedr02,tsaga13} $S(t) = -\Tr{\rho(t) \log \rho(t)}$,
of the density matrix, $\rho$,
propagated from time $t$ to time $t'$ using Eq.~\ref{e:jaynes}.
The physical content of the theory is that the classical
uncertainty in the random noise applied during time propagation
serves as an upper bound to the increase in the quantum entropy of the system,
$S(t') - S(t) \le -\sum_r P_r \log P_r$.
In addition, Jaynes showed that the unweighted process always contains the uniform
(infinite temperature) distribution as a stationary state.

  The first central question in the theory of open quantum systems is, ``How do we
assign probabilities to the collision events?''  Fermi's golden rule is the traditional
answer, but depends on assuming a directionality and time-dependence for the
external energy.
In a real interaction process, however, more collisions will occur when the central
system is moving with larger velocity.  In addition, the energy of the external system
will decrease as the collision proceeds.
The system's energy change due to the impulse will then depend
on the system's initial momentum and the external force correlation time.

  The classical nonequilibrium theory\cite{droge12} modeled these random
collisions by a maximum entropy distribution for $P(r) = |f(r)|^2$,
conditional on the average impulse size,
$\avg{(\frac{\Delta p}{\Delta t})^2}$ and the energy change,
$\avg{E' - E} = \avg{\Delta p \bar p}/m$.  Correlation (not considered in this
work for clarity) was also added by constraining the autocorrelation function,
$\avg{\frac{\Delta p(t)}{\Delta t} \frac{\Delta p(t-\tau)}{\Delta t}}$.

  This lead to the unnormalized distribution (Eq.~\ref{e:class} in an explicit form,
where $\bar p = (p(t+\Delta t)+p(t))/2$),
\begin{equation*}
P(\Delta p) \propto \exp\left\{-\Delta p^2/(2\sigma^2 \Delta t) - \frac{\beta}{2}\Delta p \bar p/m\right\}
.
\end{equation*}
This result is easily shown to be equivalent to the Langevin equation:
\begin{equation*}
p'-p = F(x)\Delta t + \Delta p = (F(x) - \frac{\beta \sigma^2}{2m} \bar p) \Delta t + \sigma d\mathcal W
.
\end{equation*}
In the quantum case, the dynamics are realized in terms of
density matrices rather than operators.
To find an analogous result, the next subsections develop this analysis
for several possible exponential biasing forms.

  Once a probability scheme is chosen, one encounters the second
central question of open quantum systems,
``Should time evolution be described by statistics of multiple individual wavefunctions or by
the statistics of the time-evolution operator?'' 
This work refrains from considering statistics among alternate density matrices or
among non-orthogonal sets of wavefunctions.

  In principle, there are two different types of prior information for assigning
probabilities to alternate transitions -- creating weights that rely on complete knowledge of the
system wavefunction, $\psi$, or incorporation of a weighting into the transition
operators themselves.  These  are dual to one another in the same sense
that the Schr\"{o}dinger picture is dual to the Heisenberg one.
Both lead to a scheme for moving from a starting pure wavefunction
to a density matrix a short instant later.
The first approach leads to transition weights,
$\avg{U_r \psi |\hat H| U_r \psi} - \avg{\psi|\hat H|\psi}$, that depend
exponentially on $\psi$.  Sec.~\ref{s:psiwt} shows this choice to be inadmissible
based on the known properties of the measurement process.
The second method fixes this defect by working in superoperator space.
Here, it is possible to average over all propagators, $U_r (\cdot) U_r^\dagger$, treating them as members
of a linear set of dimension $N^2$ (Eq.~\ref{e:canon}).
Weighting the result by performing a similarity transformation is 
equivalent to weighting each propagator, and results in a propagator
that is linear in $\rho$ except for normalization.

\subsection{ Maximum Entropy conditional on $\psi$}\label{s:psiwt}

  Since each possible time evolution operation ($U_r$) influences the
energy of the isolated system, we can choose $P_r$ in the quantum evolution
operation (Eq.~\ref{e:jaynes}) to be consistent with energy conservation.
For a prescribed average energy change, the maximum entropy form is,
\begin{equation}
P_r \propto P^0_r e^{-\tfrac{\beta}{2} \avg{\psi(0) \big| U_r^\dagger \hat H U_r - \hat H \big| \psi(0)}}
.\label{e:maxent}
\end{equation}
The probability distribution over $r$ is therefore conditional on the starting state, $\psi(0)$.
This conditioning reduces the likelihood of interactions with the
external environment that raise the energy of the system.

  Using random impulsive $U_r$ with this probability assignment, the wavefunction $\psi$ continually
diffuses through Hilbert space.  Despite the similarity to quantum state diffusion,\cite{qsd,rscha95,wstru00}
it is not clear whether any set of evolution operators and constraints will bring the two
into the same form.

  The analysis of Eq.~\ref{e:maxent} and the associated ``kinetic'' partition function,
\begin{equation*}
\Z(\lambda, \psi) = \sum_r P^0_r e^{-\lambda \avg{\psi | \Delta H | \psi}}
,
\end{equation*}
are relatively straightforward, except that the probability distribution is over the
random perturbation rather than over the physical system of interest.

  Since the propagator will preferentially weight certain outcomes over
others, it implicitly gives a model for wavefunction collapse.  Measurement
happens when one of the set of possible outcomes at time $t + \Delta t$
is selected by the environmental noise process.

  To be consistent with observation, however, the collapse
should be completely described by the density matrix
at time $t$.  Extra information is not allowed because
individual wavefunctions do not represent mutually exclusive events.\cite{jvonn55}
Rather, any two wavefunctions, say $\psi$ with $p_1$ and $\phi$ with $p_2$,
span a 2-dimensional subspace where the probability
of any measurement can be represented by arbitrary linear combinations\cite{ejayn57a}
\begin{equation}
\left[
\sqrt{p_1'} \psi'
\, \middle\vert \,
\sqrt{p_2'} \phi'
\right] \notag \\
=  \left[
\sqrt{p_1} \psi
\, \middle\vert \,
\sqrt{p_2} \phi
\right] 
 \cdot
\begin{bmatrix} \cos\theta & -\sin\theta \\
\sin\theta & \cos\theta \end{bmatrix}
.\label{e:mix}
\end{equation}

  However, the choice of Eq.~\ref{e:maxent} leads to a change in the final
density matrix ($\rho'$) when this reparameterization is done.
This dependence would make it experimentally possible to
distinguish different combinations of wavefunctions making up
the density matrix (i.e. $\theta$).  However, this is not observed
in practice.  Therefore $\psi$ is unsuitable for use as prior information
to directly determine the distribution of $U_r$.

  We note that quantum state diffusion does not seem to have this drawback,
since the form of the dynamics is such that the density matrix
found by averaging $\pure{\psi}$ at each timestep
conforms to the master equation in Lindblad form.\cite{rscha95,glind76}

  In addition to this problem, note that there is also a loss of
compactness in describing a weighting over transition processes,
$U_r$.  For example, two distinct
propagators, $U_r$ and $U_s$, could lead to the same final $\psi' = U_{r,s} \psi$.
It would then be most economical to collapse these two propagators
into a single logical possibility, reducing the upper bound of the subjective
H theorem (\ref{e:Hthm}).
  This can be done using some linear algebra to state
the transition probability in a way that does
not depend on the (arbitrary) space of random propagators, $r$.
The bare minimum description of the transition probability for
a wavefunction is given by the 4-index superoperator\cite{ejayn57a,tsaga13} (Sec.~\ref{s:math}),
\begin{equation}
G = \sum_r P_r U_r (\cdot) U_r^\dagger
. \label{e:G}
\end{equation}
Expressions built in terms of this operator will then
depend only on the actual action of the propagators on the system,
rather than the abstract space of $r$.

\subsection{ Energy-Weighted Propagator}\label{s:enwt}

  This section examines a probability assignment based directly
on a weighting of the superoperator, $G$, toward transitions
representing larger amounts of energy loss from the system.
First, re-express the transformation (Eq.~\ref{e:G})
in the energy basis as $\rho_{i'j'}' = \sum_{ij} G^0_{i'i|jj'} \rho_{ij}$.
Next, apply an exponential bias toward propagators with larger
amounts of net energy loss.  This leads to a re-weighting,
\begin{equation}
G_{i'i|jj'} = G^0_{i'i|jj'} e^{-\lambda (\Delta E_i + \Delta E_j)/2}. \label{e:rewt}
\end{equation}
The weights connecting transitions between energy eigenstates
are of maximum entropy form.
It is difficult to prove a more general statement about the entropy over transition
superoperators.  Since a more general extremum principle has not yet been found,
this approach will be termed the {\em energy-weighted propagator}
in the analysis below.

  When written in an arbitrary basis, the bias on
the energy change (Eq.~\ref{e:rewt}) has the effect of transforming each of
the unitary operators of Eq.~\ref{e:jaynes} into
$U_r \to e^{-\lambda \hat H/2} U_r e^{\lambda \hat H/2}$.
This is equivalent to transforming the $N^2$ canonical propagators
derived by an eigenvalue decomposition of $G^0$.
For proofs and related notation, see Sec.~\ref{s:math}.
Finally, we arrive at a prescription for adding a bias accounting for energy conservation
to an initial stochastic transition superoperator, $G^0$,
\begin{equation}
G = (e^{-\lambda \hat H/2}(\cdot)e^{\lambda \hat H/2}) G^0 (e^{\lambda \hat H/2}(\cdot)e^{-\lambda \hat H/2})
.\label{e:maxent2}
\end{equation}

  The description of the dynamics then becomes something like
\begin{equation*}
\tilde \rho' = G[\rho]
.
\end{equation*}
However, $\tilde \rho'$ is not necessarily normalized.
Instead, the overall normalization can be expressed as a matrix operator,
\begin{align}
\Tr{G[\rho]} &= \mathrm{Tr}_{\gets}[\sum_{ij} G_{i'i|jj'} \rho_{ij}] = \Tr{Z \rho} \notag \\
Z &= (\mathrm{Tr}_\gets[G])^T = \sum_r P^0_r e^{\lambda H/2} U_r^\dagger e^{-\lambda H} U_r e^{\lambda H/2}
. \label{e:Z}
\end{align}

  In principle, there are three options for normalizing $\tilde \rho'$.
The simplest is applying a single scaling constant to the whole
input density matrix, $G^{(1)}_N[\rho] = G[\rho] / \Tr{Z \rho}$.
Failing that, the density matrix could be scaled before applying $G$
to give a linear superoperator,
\begin{equation}
G^{(2)}_N[\rho] = G[Z^{-1/2} \rho Z^{-1/2}]. \label{e:G2}
\end{equation}
As a final alternative, the starting density matrix can be normalized by
decomposition into a set of pure states, $\rho = \sum_k p_k \pure{\psi^k}$,
each of which is given an  individual normalization constant,
\begin{align}
G_N &= \sum_k (I(\cdot)\pure{\psi_k}) G (\pure{\psi_k}(\cdot)I) / \avg{\psi_k | Z | \psi_k} \notag \\
G_N[\rho] &= \sum_k p_k G_N^{(1)}[\pure{\psi_k}] \label{e:T2}
.
\end{align}

  The first choice has nice mathematical properties, but is physically inadmissible.
It shows non-causal behavior by combining otherwise independent parts of a
statistical mixture of states.  Specifically, the single
normalization constant takes contributions from all starting states
in a mixture,
\begin{equation*}
\mathrm{Tr}_{\gets}[ G_N^{(1)} \rho ] = \sum_{k} \frac{p_k \avg{\psi_k|Z|\psi_k}}{\Tr{Z \rho}}  \Big(\pure{\psi_k}\Big) \ne \rho
.
\end{equation*}
This operator does not make a suitable transition distribution,
since summing over final states does not give back the starting probabilities,
$p_k$.  Consequently, it cannot reproduce the starting density matrix,
and will not be considered further.

  The second choice defines a linear superoperator that makes a more suitable
transition distribution, but does not act like a moment generating function
as we would expect from maximum-entropy statistical mechanics.
In particular, the first derivative of the matrix logarithm does not
give the energy change over a time-step unless $Z$ has a special form
(commutes either with the energy, $H$, or the density matrix, $\rho$).
\begin{equation*}
\Tr{-\pd{\log Z}{\lambda} \rho} = \sum_r P^0_r \Tr{ [H,U_r] Z^{-1/2} \rho Z^{-1/2} U_r^\dagger }
\end{equation*}
This equation is proved by noting the derivative,
$\partial \Tr{G_N^{(2)}[\rho]} / \partial\lambda$ is zero and
using $Z^{-1/2} = \exp(-\tfrac{1}{2} \log Z)$.
Also, the starting density matrix is not recovered using the na\"{\i}ve definition,
\begin{equation*}
[G^{(2')}_N\rho]_{i'i|jj'} = G_{i'n|mj'} Z^{-1/2}_{ni} \rho_{ij} Z^{-1/2}_{jm}
,
\end{equation*}
That leads instead to $\mathrm{Tr}_{\gets}[G^{(2')}_N\rho]_{ij} = \delta_{ij} \rho_{ij}$.
  Although it will not
be considered in this work, numerical experiments
(not shown) find that this operator behaves much like $G_N$,
when the noise level in $P_r^0$ is much smaller.

  The third choice propagates orthogonal wavefunctions belonging to the initial density matrix separately.
This leads to a proper causal picture\cite{ejayn57a} where a single pure
state chosen from a statistical mixture at each point in time is
responsible for the system's future behavior.
It reduces to the second choice, $G^{(2)}_N$, when $Z$ commutes with $\rho$.

  It is also simple to check that propagating the mixture of Eq.~\ref{e:mix}
gives an answer independent from $\theta$,
\begin{align*}
\rho' &= p_1' G_N[\pure{\psi'}] + p_2' G_N[\pure{\phi'}] \\
 &= (p_1' \norm{\avg{\psi | \psi'}}^2 + p_2'\norm{\avg{\psi | \phi'}}^2) G_N[\pure{\psi}] \\ 
  &\quad + (p_1'\norm{\avg{\phi | \psi'}}^2 + p_2' \norm{\avg{\phi | \phi'}}^2) G_N[\pure{\phi}] \\
&= p_1 G_N[\pure{\psi}] + p_2 G_N[\pure{\phi}]
.
\end{align*}
Interestingly, this means that, although $G$ must be specialized for $\rho$,
the choice of propagating only the eigenstates of $\rho$ in Eq.~\ref{e:T2} is not unique.
The same final density matrix, $\rho'$, would result from propagating any
non-orthogonal weighted set of starting states which combine
to give $\rho$ (an array in the terminology of Ref.~\cite{ejayn57a}).

\subsubsection{ Transition Free Energy and Entropy}

  The remainder of the paper introduces a reference process and
proceeds to analyze the qualitative and analytical mathematical
properties of the third choice, $G_N$.
In particular, we find that a kinetic free energy functional,
\begin{equation}
F = -\sum_k p_k \log{\avg{\psi_k|Z|\psi_k}}
, \label{e:F}
\end{equation}
(where $Z$ was defined in Eq.~\ref{e:Z})
has properties consistent with maximum entropy thermodynamics.

  A nonnegative relative information entropy can be defined for each transition,
\begin{equation}
dS_\text{tot} = \Tr{ G_N \rho \left(\log G_N \rho - \log \rho' G_N'\right)}
\label{e:S}.
\end{equation}
By the Klein inequality, this is necessarily positive when $\rho' G_N'$
is any normalized, completely positive 4-index tensor.

  Conventionally, $G_N'$ is chosen to be a ``reverse'' transformation from
the final density $\rho'$ to an inferred joint density matrix -- so that
$dS_\text{tot}$ is a measure of the irreversibility of the process during time $\Delta t$.\cite{gcroo99,cjarz08}
This choice appeals to information theory on the grounds that,
after discarding $\rho$ in favor of the updated $\rho'$, $G_N'$ presents a
``best guess'' as to the 2-time probability density of the system.\cite{droge11a}
Since $dS_\text{tot}$ is the information lost during the update
step, it must obviously go to zero for deterministic evolution and depend on the ratio
of forward to reverse probabilities.

  Physically, Eq.~\ref{e:S} should be derived by considering the trade-off between
spontaneous decrease in the system's information entropy and
heat released into the environment.\cite{droge12}  In this case, the reason
for using a reverse transformation is so that $G'$ depends on temperature like
$e^{\beta\Delta E/2}$.  The difference, $\log G - \log G'$, then contains a
heat release term (negative of the system's energy gain $\avg{\beta \Delta E}$)
that counter-balances increase or decrease of the system's
instantaneous information entropy, $\Delta S_\text{inf} - \avg{\beta \Delta E} \sim \Delta S_\text{tot} \ge 0$.
This should be compared to the classical result (Eq.~\ref{e:Sclas}), where, for quantum
systems,
\begin{equation*}
S_\text{inf} = -\Tr{\rho \log \rho}
\end{equation*}
Re-stated in terms of the system itself, the system cannot store thermal energy
without increasing its information entropy.

  A straightforward choice based on these considerations is,
\begin{equation*}
G' = (e^{\lambda/2 \hat H}(\cdot)e^{-\lambda/2 \hat H})
G^0 (e^{-\lambda/2 \hat H}(\cdot)e^{\lambda/2 \hat H})
,
\end{equation*}
and consequently
\begin{equation*}
G_N' = \sum_k (\pure{\psi_k'}(\cdot)I) G' (I(\cdot)\pure{\psi_k'}) / \avg{\psi_k'|G'[I]|\psi_k'}.
\end{equation*}
Note that the direction of time plays no special role other than swapping the
application order of $\rho'$ -- matching the final indices $i',j'$ so it
encounters $U^\dagger(\cdot)U$.
The results in Sec.~\ref{s:collapse} show that this definition
is most closely related to the information loss interpretation,
since it decomposes approximately as,
\begin{equation*}
dS_\text{tot} = dS_\text{inf} + \beta dQ_\text{irrev}
.
\end{equation*}

\subsection{ Reference Process}\label{s:ref}

  To illustrate the features of the energy-weighted propagator, it is helpful to compare
against a reference process.
Many recent experiments on molecular beams and trapped atomic systems
have been carried out to probe the dynamic properties of wavefunction collapse.\cite{csun97,qturc00,qturc00a,lhack04}
Although experiment makes the best reference in principle,
direct comparison is limited by uncertainties in the source and
strength of  environmental noise.  These differences, reflected in alternative choices
for $U_r$, make noticeable differences as the noise strength increases.\cite{qturc00a}
In addition, the Hamiltonians for most experimental setups introduce
additional complexities which are not essential for understanding the theory.

  We therefore consider a physically motivated theoretical reference
process for environment-induced decoherence and dissipation.
This provides a strong intuition for the role of energy-weighting
in the stochastic propagator of the density matrix.

  The simplified reference system consists of two resonant harmonic
oscillators.\cite{gford87}  Oscillator $a$ represents the system of interest (e.g. the atom in a Paul trap).
Oscillator $b$ represents a single electromagnetic mode of an
enclosing cavity.  Spin is ignored, and the two are simply coupled by an energy term, $c x_a x_b$.
The Hamiltonian is thus:
\begin{equation*}
\hat H = \hbar \omega (\hat a^\dagger \hat a + \hat b^\dagger \hat b + 1) + c \frac{\hbar}{2m\omega} (\hat a^\dagger + \hat a) (\hat b^\dagger + \hat b)
,
\end{equation*}
where $\hat a,\hat b$ are the lowering operators for the uncoupled harmonic oscillators.
Similar models using a harmonic oscillator as the environment and a
2-level atom as the system of interest have also been investigated in the context
of deriving fluctuation theorems for exact dynamics.\cite{mcamp09,mcamp09a}
Further elaborations on this model to include the atomic spin and the direction
and polarization of the field\cite{wheit36,LL4,rtoug79} could be used for
comparison to experiment.\cite{qturc00a}

  The environment in our reference model has the responsibility for introducing
both decoherence and energy dissipation.  To model this, we propose a
physically motivated thermostatting scheme.  It is based on the Andersen
thermostat for molecular dynamics.  There, the velocity of a particle is periodically
replaced with a random sample from the Boltzmann distribution.\cite{hande80}
By analogy, this scheme periodically replaces the cavity state with a pure energy
level sampled from the Boltzmann distribution.

\begin{figure}
\includegraphics[width=0.45\textwidth]{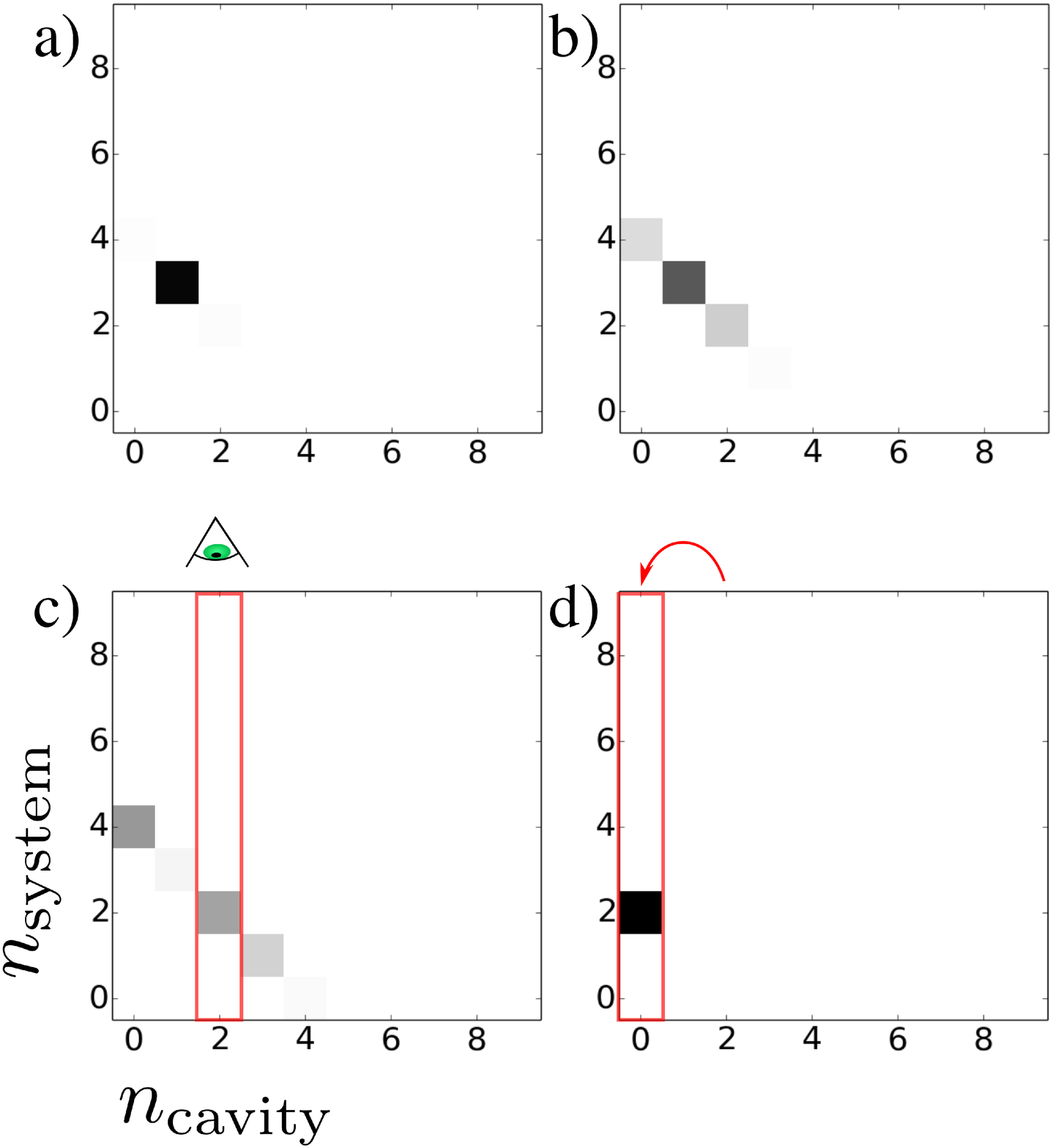}
\caption{Schematic of the quantum Andersen thermostat (reference process).
(a-c) Exact evolution of the coupled system+cavity from a Fock state
quickly leads to an entangled state.
(c) Measuring the cavity energy selects a subsample of the system,
removing coherences.
(d) Replacing the cavity state with a thermal sample adds or removes
energy.  The thermal nature of the environment is responsible for dissipation.}\label{f:ref}
\end{figure}

  The process can be visualized as in Fig.~\ref{f:ref}.
The joint wavefunction is represented in the uncoupled energy space as
\begin{equation*}
|\psi\rangle(t) = \sum_{n,m=0} \psi_{nm}(t) |n\rangle_a \otimes |m\rangle_b
.
\end{equation*}
The panels of Fig.~\ref{f:ref} show probability
densities for these joint states, $|\psi_{nm}|^2$.
Starting from a state where both energy levels are known in \ref{f:ref}a,
the joint system undergoes deterministic evolution according to
\begin{equation*}
|\psi\rangle(t) = e^{-\tfrac{it}{\hbar} \hat H} |\psi\rangle(0),
\end{equation*}
which introduces a system-field superposition (\ref{f:ref}b-c).
Measurement of the cavity mode (\ref{f:ref}c) collapses the system
into one of the pure cavity states,
$\psi_a \otimes |m\rangle_b = (\sum_{n} \psi_{nm}(t) |n\rangle_a ) \otimes |m\rangle_b$,
chosen in the usual way with probability $P(m) = \sum_{n} |\psi_{nm}|^2$.
Finally, the cavity mode is replaced (\ref{f:ref}d) with a new state sampled from the
thermal distribution,
\begin{equation*}
\psi_a \otimes |m\rangle_b \to \psi_a \otimes |m'\rangle_b, \quad
P(m') \propto e^{-\beta\hbar\omega m'} \notag
.
\end{equation*}

  There is a distinction in this process between decoherence and dissipation.
Decoherence occurs when the environmental state is measured, since
a state containing all the information about the the system-cavity superposition is
projected onto a single cavity mode.  This step usually introduces small increases in the energy
of the joint system because the coupling, $c x_a x_b$
has only off-diagonal terms in $m$ -- which are removed by the
projection.
However, replacement of the cavity mode
with a thermal sample accomplishes energy exchange -- heating or cooling
the system until thermal equilibrium is obtained. 

\subsubsection{ Deterministic Limit}

  The measurement process described above incessantly projects
the environment into a pure state.  This type of measurement
traces over the environmental degrees of freedom, so that
at each measurement point the environment is seen to follow a deterministic path
without superpositions.  Starting with random off-diagonal matrix elements
for the environment will have the same effect -- canceling paths that
begin in a superposition state.

  If the trace (measurement) process is made continuous, the environment follows
a definite trajectory, and is only felt as a deterministic external force --
replacing $\hat H_{ab}$ with $\hat H_a + V(x_a,x_b)$.
This is the quasi-classical model used for many time-dependent
quantum processes, where the Hamiltonian depends predictably on time
through $x_b$.

\section{ Methods - Superoperator Notation}\label{s:math}

  For analyzing the re-weighting of Eq.~\ref{e:rewt}, we introduce the following
notation for density matrix superoperators.
The canonical form for a superoperator is\cite{vvedr02}
\begin{equation}
G = \sum_{k=1}^{O(N^2)} \gamma_k C_k (\cdot) C_k^\dagger \label{e:canon}
.
\end{equation}
It is a 4-index quantity.  Explicitly, $G_{i'i | jj'} = \sum_k \gamma_k C_{k,i'i} C^*_{k,j'j}$.
Unless otherwise noted, each index (i,j,{\em etc.}) ranges over integer energy levels, $0,\cdots,N-1$.
Superoperators appropriate for the density matrix can all be written in
the form of Eq.~\ref{e:canon} by virtue of the (super-Hermitian) symmetry property
$G_{i'i|jj'} = G^*_{j'j | ii'}$.  This property also guarantees a spectral
decomposition into $N^2$ matrices, $\{C_k\}$, which are mutually orthogonal -- as
defined by $\sum_{i'i} C_{k,i'i} C^*_{l,i'i} = \Tr{C_l^\dagger C_k} = \delta_{kl}$.

  We define the inner and outer trace operations,
\begin{align*}
\mathrm{Tr}_\to [G]_{i'j'} &= \sum_{ij} T_{i'i|jj'} \\
\mathrm{Tr}_\gets [G]_{ij} &= \sum_{i'j'} T_{i'i|jj'}
.
\end{align*}

  Superoperators are applied to matrices by a substitution pattern,
\begin{equation*}
(A(\cdot) B) [\rho] = A \rho B,
\end{equation*}
and are composed following the same pattern,
\begin{align*}
(A(\cdot) B) \circ (C(\cdot)D) &= (A(\cdot) B) [ (C(\cdot)D) [\rho] ] \\ 
 &= (AC(\cdot) DB)[\rho]
.
\end{align*}
We also define a simple algebraic notation for multiplication,
\begin{equation*}
(A(\cdot) B) (C(\cdot)D) = (AC(\cdot) BD)
\end{equation*}
so that
\begin{equation*}
\tens{A}{B} \circ \tens{C}{D} = \tens{A}{I} \tens{C}{D} \tens{I}{B}
.
\end{equation*}

  Note the difference in application order between composition and multiplication.
The multiplication operation makes certain manipulations clearer.
In this notation, the Choi-Jamiolkowski isomorphism is simply obtained by writing
the identity for a complete basis,
\begin{equation*}
G = \sum_{ij} \tens{I}{\pure{j}} G \tens{\pure{i}}{I}
.
\end{equation*}
Writing an initial transition tensor explicitly in the energy basis, $\{|k\rangle\}_0^{N-1}$,
\begin{equation*}
G^0 = \sum_{i'i,j'j} G^0_{i'i|jj'} \tens{|i'\rangle\langle i|}{|j\rangle\langle j'|}, \label{e:Texpl}
\end{equation*}

  We define the direct product of a superoperator and a density matrix, $\rho$,
as the projection onto the eigenspace of the density matrix ($\{p_k,\psi_k\}_0^{N-1}$),
\begin{equation*}
G\rho = \sum_k p_k (I(\cdot)\pure{\psi_k|}) G (\pure{\psi_k}(\cdot)I)
,
\end{equation*}
so that $G[\rho] = \mathrm{Tr}_\to [G \rho]$.  This operation effectively
specializes a superoperator to be consistent with its operand, $\rho$.
The choice to project or not does not affect the discussion
in Sec.~\ref{s:deriv} leading to $G_N$.

  The numerical results presented here made use of tabulated $G^0_{i'i|jj'}$
tensors calculated by numerically integrating Eq.~\ref{e:noise}
for $|f(r)|^2| = \exp(-r^2/2\sigma^2)/\sqrt{2\pi\sigma^2}$
using a 20-point Gauss-Hermite quadrature rule
and an exact matrix exponential computed through
eigenvalue decomposition.

\section{ Results}\label{s:results}

  The first role of the environment is to provide a mechanism for decoherence by
introducing random noise into the time evolution.
This section presents results from the reference process,
which show the importance of the total energy in determining
intermediate states just before collapse.

\subsection{ Dynamics of the Reference Process}

\begin{figure*}
\includegraphics[width=0.9\textwidth]{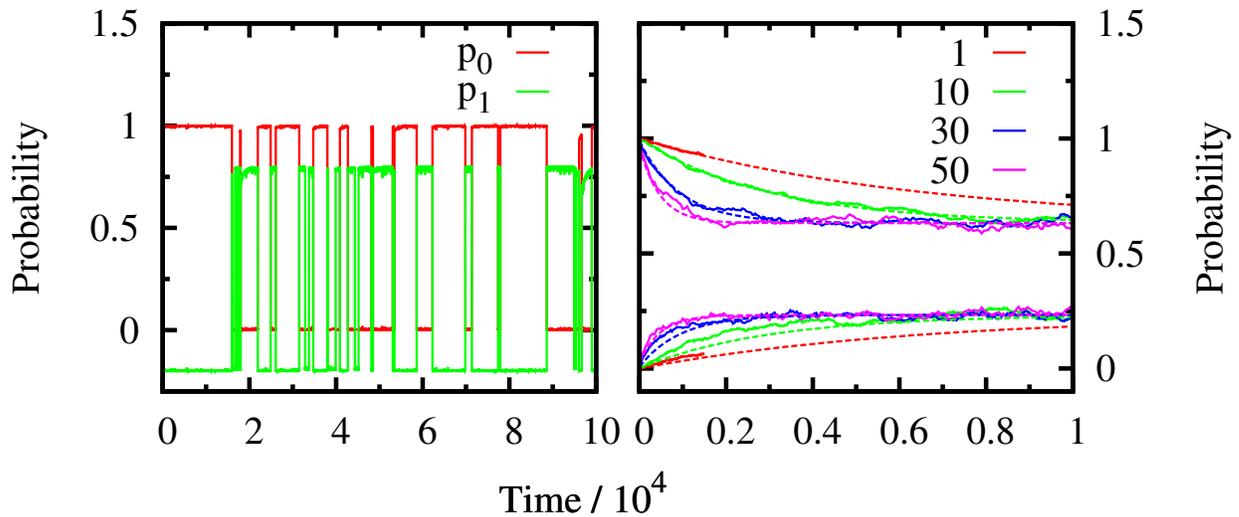}
\caption{Panel (a) shows occupation probabilities for the ground and first excited state of a central system interacting with a cavity under quantum Andersen thermostatting ($\hbar\omega=1,\beta=0.1,c=0.01,t_\text{reset}=10$).  The $p_1$ curve is shifted down by 0.2 for visual clarity.
Panel (b) shows the time-course of the average occupation probability for ground ($p_0$, upper curves) and first excited states ($p_1$, lower curves).  The parameters are the same as panel (a), except $\beta = 1$.  Solid lines are averages over 1000 trajectories.  Dashed lines show exponential relaxation with a rate constant fit from $p_0(t)$.}
\label{f:ref_traj}
\end{figure*}

  The reference process makes quantum jumps through periodically
projecting the cavity into a Fock state.  Although this always leaves the central
system in a superposition state, Fig.~\ref{f:ref_traj} shows that
the probabilities for each system state still appear to follow a jump process.
The resulting trajectory is very reminiscent of fluorescence measurements
of single-atom excited state probabilities.\cite{wnago86}

  We find numerically that the Andersen-type thermostat scheme
employed for the reference process leaves the system very nearly in
the equilibrium state of the coupled system.
However, when the coupling strength is large, projecting out the cavity
from the combined equilibrium density matrix
gives a system density that can deviate significantly from the Boltzmann
distribution (Fig.~\ref{f:equ}).
Practically, this means that (for physically correct reasons),
the thermostat does not work at strong coupling and only yields
the exact Boltzmann distribution in the zero-coupling limit.
We shall find similar behavior for the maximum-entropy jump process.

  We first determine the dependence of the decorrelation time and energy dissipation in the
reference process as a function of the coupling strength, $c$, and the time between
interactions, $t_\text{reset}$.  Fig.~\ref{f:ref_traj} shows the average system energy level
vs. time over 1000 stochastic trajectories, each starting from the ground state.
The index labels the system energy level, projected from the complete system plus cavity state.
Each parameter set was run for a length of 1500$\times t_\text{reset}$,
using $N=10$ energy levels for each system and a numerical matrix exponent
between each reset.\cite{cmola03}
The time-course of $p_0$ fits well to an exponential with relaxation
rate that increases approximately as $t_\text{reset}^{3/2}$.

  We have verified numerically that the relaxation rate depends
on the square of the coupling strength, $c$, as predicted by
perturbation theory.\cite{wzure02}
Finally, a dependence on temperature will come from the dependence of
the jump probability on energy level.
Since the jump probability between $n,m+1$ and $n+1,m$ states by perturbation theory
is proportional to $nm$, we might expect quadratic dependence on $\avg{m}+1/2$.
However, at high temperature the jump probability empirically shows a linear,
rather than quadratic dependence on $\avg{n}$.
Fitting the observed relaxation rates, $r$, using this scaling resulted in reasonable
agreement for the constants,
\begin{equation}
r \simeq 0.0510747 \frac{c^2 \avg{m+1/2}}{\omega} (\omega t_\text{reset})^{3/2}
+ 0.000154756 \label{e:fit}
.
\end{equation}

\subsection{ Properties of the Energy-Weighted Propagator}

  This section begins by showing that the qualitative dynamics and approach to equilibrium
of the energy weighted propagator is very similar to the reference process above.
Next, simple examples of the kinetic partition function are given.  The section concludes
by showing the relation between work, heat and entropy production.

\begin{figure*}
\includegraphics[width=0.9\textwidth]{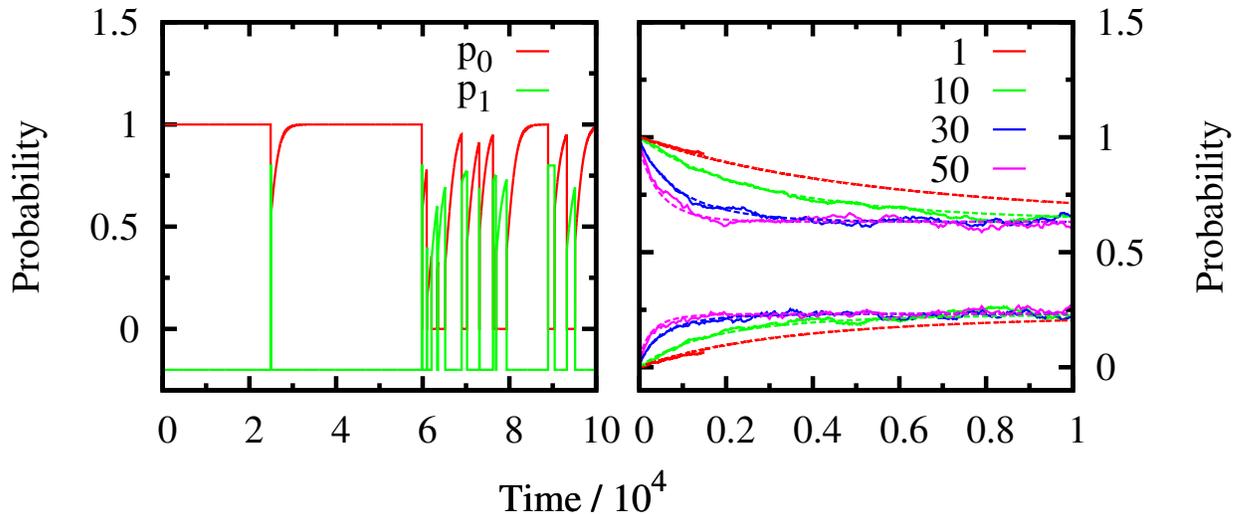}
\caption{Dynamics of the weighted energy exchange propagator.
(a) Single trajectory ($\beta = 0.1$) integrated by choosing a pure state from the density
matrix at each time point.  The $p_1(t)$ curve has been shifted down by $0.2$ for visual
clarity.
(b) Comparison of combined density matrix evolution (smooth curves) with analogous
quantum Andersen thermostat result from Fig.~\ref{f:ref_traj}b.
The temperature, $\beta = 1$, translates to $\lambda = 1/2$ (Sec.~\ref{s:equ}).}
\label{f:wt_traj}
\end{figure*}

  Fig.~\ref{f:wt_traj} shows the time-dependence
of a single-harmonic oscillator simulated with the maximum-entropy
matrix propagator derived in Sec.~\ref{s:enwt}.
The variance of the random impulsive forces was chosen
to give the relaxation rate corresponding to the reference
process (using Eq.~\ref{e:fit}).
The variance of momentum jumps is related to the diffusion
constant of Langevin dynamics as $\sigma^2 = 2D\Delta t$.\cite{droge12}
As should be expected, the energy relaxation rate was found to be
linear in the diffusion constant, at $r = 1.42397 D$ per step.

  The dynamics of the weighted propagator appears qualitatively similar to the reference
process.  Fig.~\ref{f:wt_traj}a shows jump behavior that
indicates the propagator induces localization of the wavefunction in energy space.
The transitions are slightly more random, as might be expected from a
maximum uncertainty model for interaction with the environment.
The average probabilities in panel \ref{f:wt_traj}b (dashed lines) fit the reference process
(solid lines) noticeably better than a simple exponential
-- compare to the green line in Fig.~\ref{f:ref_traj}.
We stress that the only fitting parameter is the diffusion constant, which was chosen
so that the relaxation times agree between the two processes.

\subsubsection{ Stationary Density Matrix}\label{s:equ}

  The propagator $G_N^{(1)}$ admits the Boltzmann-Gibbs distribution with
$\beta = \lambda$ as a stationary state of the dynamics.
This is shown since $G[e^{-\lambda H}] = e^{-\lambda H/2} G^0[I] e^{-\lambda H/2} = c e^{-\lambda H}$,
and $G^0[I] = I$.  The unweighted propagator always preserves the infinite
temperature distribution, since it is an average over
unitary operations (Eq.~\ref{e:jaynes}).

\begin{figure*}
\includegraphics[width=0.9\textwidth]{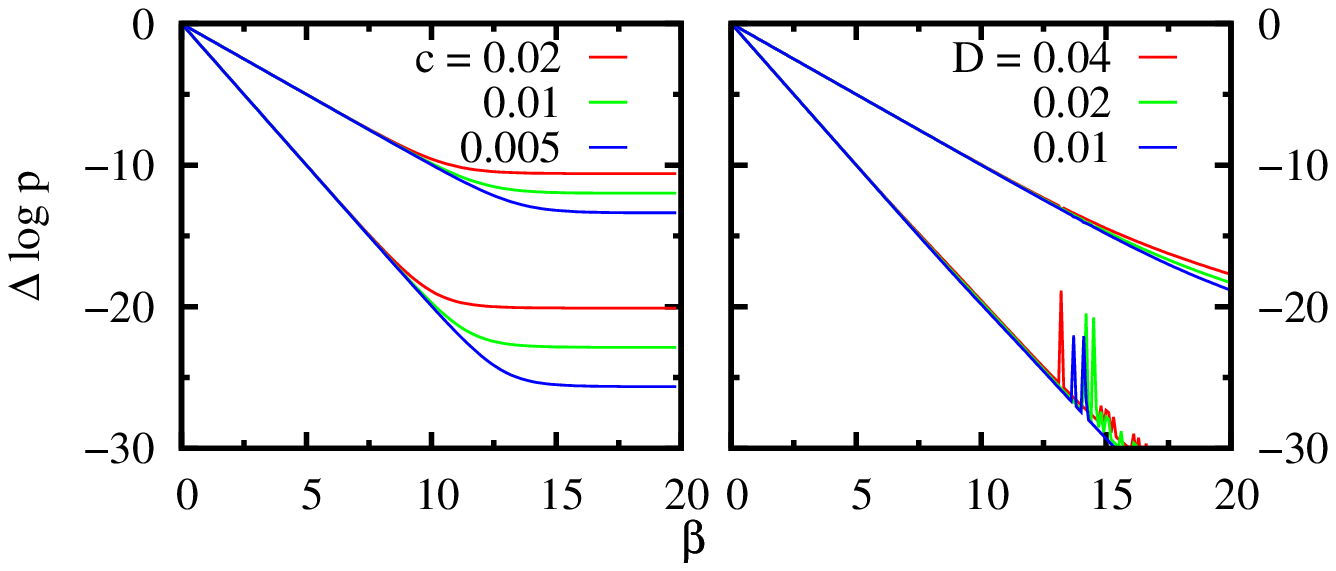} \\
\includegraphics[width=0.9\textwidth]{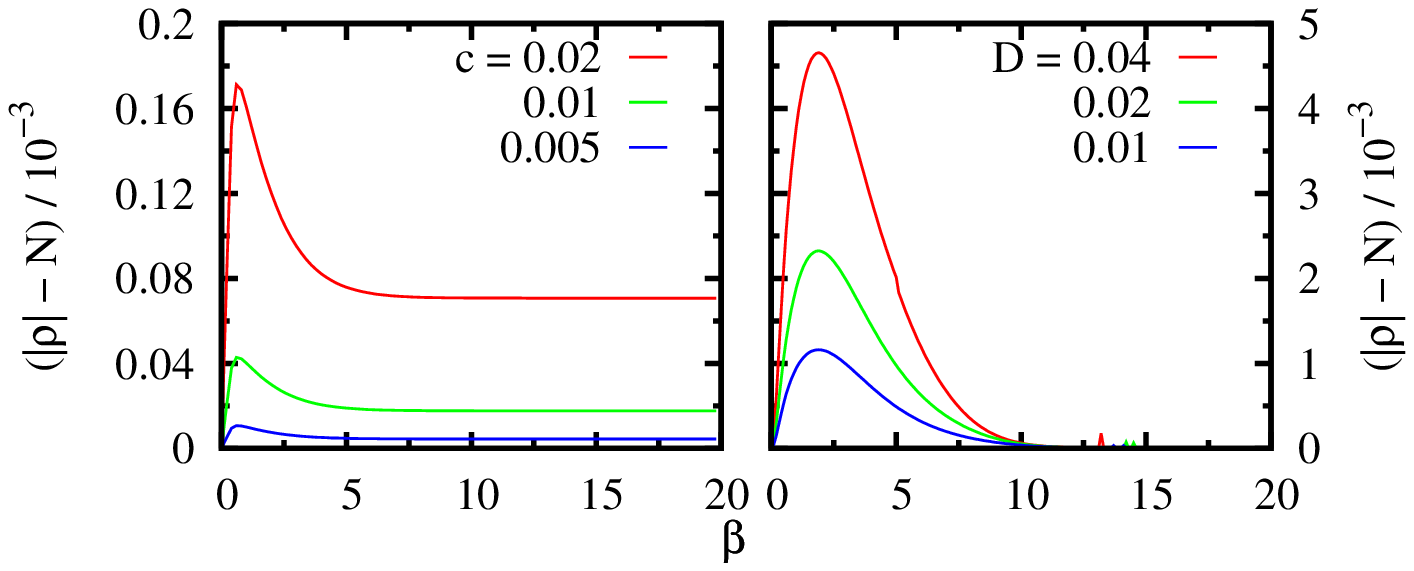}
\caption{Temperature-dependence of the stationary states of (a,c) reference
Hamiltonian and (b,d) energy-weighted propagator.
Each color corresponds to a different coupling parameter (in a,c)
or diffusion constant (in b,d).
For (a) and (b), the upper set of lines show the first excited state log ratio, $\log p_1/p_0$,
while the lower set show the second excited state ratio, $\log p_2/p_0$.
The Boltzmann-Gibbs distribution would show straight lines with slope -1 and -2, respectively.
Numerical noise at low temperature in (b) is due to difficulty converging the very low
probability of the second excited state, $p_2$.
Panels (c) and (d) show the total magnitude of the off-diagonal
elements of the $10\times 10$ density matrix.}\label{f:equ}
\end{figure*}

  The propagator $G_N$ (Eq.~\ref{e:T2}) would not normally be expected to exhibit a
Boltzmann-Gibbs stationary distribution.
However, empirically the stationary state of $G_N$ at small $\lambda$
is very close to the Boltzmann-Gibbs distribution with $\beta = 2\lambda$ (Fig.~\ref{f:equ}).
At large $\lambda$, the stationary state becomes more like the Boltzmann-Gibbs
distribution with $\beta=\lambda$ for $G^{(1)}$.

  The stationary distributions in Fig.~\ref{f:equ} were found by repeated eigenvalue
analysis on the re-normalized propagator, $G_N$.  Convergence was achieved
when the eigenvectors of $\rho$ rotated with $N - \sum_k |\avg{\psi_k'|\psi_k}| < 10^{-4}$.
The stationary state shows two regimes.  The high-temperature,
low diffusion constant limit gives a density matrix approximately proportional to $e^{-2\lambda H}$.
The low-temperature, high diffusion constant limit gives instead the density matrix $e^{-\lambda H}$.
This behavior is evident at the break in the slope of probability vs. $\lambda$.

  This behavior can be justified when there are enough energy levels in the
system such that $G^0_{(m-n)m|m(m-n)} = G^0_{(m+n)m|m(m+n)}$,
with $G^0_{(m+n)m|m(m+n)}$ decaying with increasing energy difference, $n$.
In this case, the moment generating functions,
\begin{equation*}
W_m(\lambda) = \sum_n G^0_{nm|mn} e^{-\lambda(E_n - E_m)}
\end{equation*}
are nearly independent of the starting state, $n$, for small $\lambda$.  If they are independent,
the Boltzmann-Gibbs distribution obeys the detailed balance condition,
\begin{equation*}
e^{-2\lambda E_n} \frac{G^0_{mn|nm} e^{-\lambda(E_m-E_n)}}{W_n(\lambda)}
  = e^{-2\lambda E_m} \frac{G^0_{nm|mn} e^{-\lambda(E_n-E_m)}}{W_m(\lambda)}
  ,
\end{equation*}
so the equilibrium probability of energy level $m$ is $\pi_m \propto e^{-2\lambda E_m}$.

  As the coupling strength ($\lambda$) is increased, the moment generating functions tend to
their extreme value, $W_m(\lambda) \to G^0_{0m|m0} e^{\lambda E_m - \lambda E_0}$.  Detailed balance then gives $\pi_m \propto e^{-\lambda E_m - \log G^0_{0m|m0}}$.  The probability of jumping to the lowest energy level, $G_{0m|m0}$, depends on the collision rate parameter.  As the rate parameter decreases, the cross-over point where the effective temperature of the stationary state doubles (from $\beta = 2\lambda \to \lambda$) is delayed to ever lower temperatures.  This behavior is seen clearly in the numerical results.

  The upper left panel of Fig.~\ref{f:equ} shows that an analogous effect happens in the reference model.  There, the thermal state with the coupled system-cavity Hamiltonian depends on the temperature and coupling strength.  When the coupling strength is small, tracing over the environment gives very nearly the Boltzmann-Gibbs density matrix for the system.  As the coupling strength increases, the system's projected density matrix
plateaus to a constant, minimum, temperature at smaller $\beta$.  This may explain
the requirement for a rescaled effective temperature in interpreting electron transport experiments
probing the fluctuation theorems.\cite{yutsu10}

  The lower panels of Fig.~\ref{f:equ} show the total magnitude of all off-diagonal
elements in the density matrix as a function of temperature.  Although these are on the order
of $10^{-3}$ ($N=10$), both the reference process and the energy-weighted
propagator show a maximum at intermediate temperatures.  In the reference system,
the off-diagonal elements are mainly two above the diagonal, and can be traced
to the double-excitation process, $a^\dagger b^\dagger$,
in the linear coupling, $x_a x_b$.

\subsubsection{ Moments of the Flux Distribution}\label{s:flux}

  The function $F(-\beta,\rho)$ (Eq.~\ref{e:F}) acts as a cumulant generating function for the distribution  of the energy flux.  This is easy to show using successive differentiation of
$\Tr{Z\rho}$.  From the definition of $G$ (Eq.~\ref{e:maxent2}),
\begin{align*}
\pd{^n \Tr{Z\rho}}{(-\lambda)^n} &= \frac{1}{2^n} \sum_{\mathfrak C} \frac{n!}{i!j!k!l!} \operatorname{Tr}\Big[ (H^i(\cdot)(-H)^j) \\
         & \qquad\qquad G ((-H)^k(\cdot)H^l) [\rho] \Big] \\
 &= \sum_{i=0}^n \binom{n}{i} \Tr{ H^i G[(-H)^{n-i} \rho] }
.
\end{align*}
The summation set, $\mathfrak C$, contains nonnegative integers, $i,j,k,l$ that sum to $n$.

\begin{figure}
\includegraphics[angle=-90,width=0.45\textwidth]{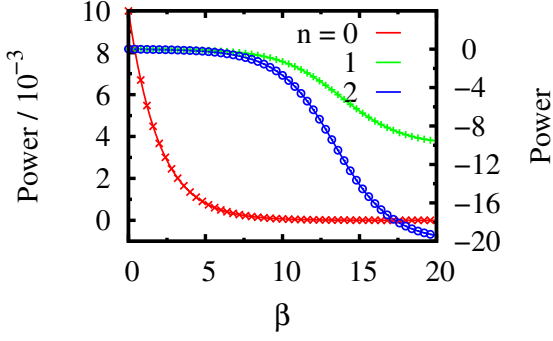}
\caption{ Energy change per unit time for a system initially in an energy eigenstate.  The ground state ($n=0$) absorbs a small amount of energy, and is plotted separately on the left scale.  At small $\beta$, other states (plotted on the right scale) initially absorb small amounts of energy but eventually lose energy quickly at lower temperature.  Lines give the result from differentiating $F$ (Eq.~\ref{e:F}), while the points show a separate calculation
as the average energy change from a single time-step of dynamics.}\label{f:work}
\end{figure}

  The cumulants of the energy flux (derivatives of $F$) therefore break into
sums over cumulants from each starting state, $a$,
\begin{align}
\pd{^n F}{\lambda^n} &= \sum_a p_a (-1)^n K^n_j
\end{align}
While each cumulant, $K^n_j$ is defined in the usual way in terms of $n^\text{th}$ order
polynomial expressions in the moments,
\begin{align*}
K^1_j &= \Tr{\hat H G_N[\pure{\psi_j}] } - \avg{\psi_j | \hat H | \psi_j} \\
K^2_j &= \Tr{\hat H^2 G_N[\pure{\psi_j}] } + \avg{\psi_j | \hat H^2 | \psi_j} \notag \\
 & - 2 \Tr{\hat H G_N[\hat H \pure{\psi_j}] } - (K^1_j)^2 , \\
  \; \text{\em etc.}
\end{align*}
These are a suitable definitions for the $n^\text{th}$ order cumulants of the distribution of
$\Delta H$.\cite{mcamp11}  When $\rho$ and $\rho'$ are both energy eigenstates, it reduces to the classical statistical distribution~\cite{cgabr15}.

\subsubsection{ Wavefunction Collapse}\label{s:collapse}

\begin{figure*}
\includegraphics[width=0.9\textwidth]{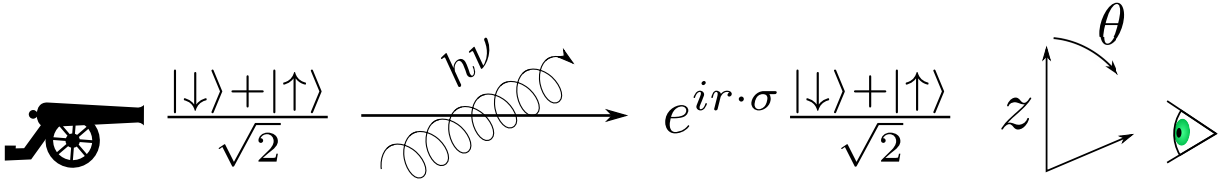}
\caption{ A schematic drawing of a molecular beam experiment
where an entangled state ($x$-spin) is subjected to a random perturbing field
before being detected at angle $\theta$.}\label{f:interfere}
\end{figure*}

  To illustrate the properties of the entropy, we give details on the process of wavefunction collapse with the energy-weighted propagator.  This will be illustrated with a simple molecular
beam experiment (Fig.~\ref{f:interfere}).\cite{beam}  Here a stream of atoms in a maximally entangled
spin state are subjected to a random perturbing magnetic field during
flight from a source to a detector.  The final density matrix can be measured by
counting the number of spin-up particles observed along each direction.

  A uniform external magnetic field provides the energy function needed to induce
wavefunction collapse.  Conventionally, the field is said to `measure' the particle's
spin state.  Because of the field, the particles arriving in the detector are
only found in one of two states - up or down along the $z$-axis.  This investigation
of thermalization therefore demonstrates the hallmark of wavefunction collapse
-- that measurement forces the system onto the diagonal in the measurement basis.

  The mathematical properties of this system are similar to recent
experimental investigations on decoherence rate of quantum systems
achieved using Ramsey interferometry on
trapped atoms in superpositions of coherent harmonic oscillator states.\cite{qturc00a}
There, each of two states is tagged with a spin.  A random displacement shifts the
relative phase of the two states, resulting in decoherence that can
be probed by moving the states back together spatially and then
measuring the probability of up-spin along each detector angle, $\theta$.

  The propagator corresponding to evolution of a spin in an external magnetic
field is $U_r = e^{i r\cdot\sigma}$, where $\vec \sigma$ represents a vector of
Pauli spin matrices, and the direction is proportional to the magnetic field,
$\vec r = \frac{\Delta t}{\hbar} \frac{g_s \mu_\nu}{\hbar} \vec B$.
The constants $g_s \mu_\nu$ stand for the appropriate gyromagnetic
ratio and magneton.\cite{Bes}
Its operation can be computed exactly using a general
formula for the exponential of a trace-free $2\times 2$ matrix,\cite{cmola03}
\begin{equation}
e^{i R} = \cos{|R|}  I + i\frac{\sin{|R|}}{|R|} R
,
\end{equation}
where $R = r\cdot\sigma = r_x\sigma_x + r_y\sigma_y + r_z\sigma_z$
and $|R| = \sqrt{r_x^2+r_y^2+r_z^2}$ is the magnitude of both
(positive and negative) eigenvalues of $R$.
The action of the unitary operator can be calculated on a $2\times 2$ Hermitian
matrix by expanding as $\rho = a I + C$, where $C = c\cdot\sigma$,
\begin{align}
e^{iR} (a I + C) e^{-iR} &= \rho + [e^{iR}, C] e^{-iR} \\
 &= \rho + \frac{i \sin{|R|}\cos{|R|}}{|R|} [R, C] \notag \\
   & + 2\frac{\sin^2{|R|}}{|R|^2} \left(c\cdot r R - |R|^2 C \right)
\end{align}

  Its average effect on the density matrix is determined by the
statistics of the direction.  Integration over $r$ is difficult in general,
but we can derive useful results by choosing
$r_x = r_{xy} \cos\phi, r_y = r_{xy} \sin\phi$ with fixed $r_z \equiv \alpha |R|$,
$|R|^2 = r_{xy}^2 + r_z^2$.
Averaging over a uniform distribution for $\phi$ gives
an unnormalized propagator,
\begin{align}
\tilde\rho' &= \rho + \alpha \sin(2|R|)
\begin{bmatrix} & c_y + i c_x \\
c_y - i c_x &  \end{bmatrix} \\
&\quad + \sin^2{|R|}
\begin{bmatrix}
2(\alpha^2 - 1) c_z & -(\alpha^2+1)(c_x - i c_y) \label{e:rhop} \\
-(\alpha^2+1)(c_x + i c_y) & -2(\alpha^2 - 1) c_z
\end{bmatrix} \notag
.
\end{align}
When $\alpha = 1$, $r_{xy} = 0$, so there are no random forces.  In this case,
the density matrix above reduces to the exact equation of motion for a
particle precessing in a magnetic field along the $z$-direction.

 The temperature-dependence of the stationary state agrees with the
qualitative results on the harmonic
oscillator from Sec.~\ref{s:equ}.  It resembles
$e^{-\beta H}$, while deviating toward larger effective
temperatures for large $\beta$ when the noise is strong
($\alpha << 1$, not shown).

  Extracting the matrix elements from Eq.~\ref{e:rhop} leads to the
transition tensor,
\begin{align}
G &= 
\upM{\begin{matrix} 1 - a & \\ & ae^{\beta E} \end{matrix}}
{\begin{matrix} & 1 - b \\ & \end{matrix}}
{\begin{matrix} & \\ 1 - b^* & \end{matrix}}
{\begin{matrix} ae^{-\beta E} & \\ & 1 - a \end{matrix}} \\
a &\equiv (1 - \alpha^2) \sin^2{|R|} \notag \\
b &\equiv (1 + \alpha^2) \sin^2{|R|} - i \alpha \sin(2 |R|) \notag
\end{align}
The outer indices are $i',j'$ for $\rho'$, with each $2\times 2$ sub-block
indexed by $i,j$ for $\rho$.

  The expectation of the energy change per unit time can
be found from $\avg{dH} = -\pd{\log\Tr{Z \rho}}{\beta/2}$.
For a state with $P(\downarrow) = P(\uparrow) = \tfrac{1}{2}$,
random collisions cause the energy to decrease
on average,
\begin{align*}
\avg{dH} &= -\pd{}{\beta/2} \log (1-a + a\cosh(\beta E) ) \\
&= \frac{ -2 a E \sinh(\beta E)}{1-a + a\cosh(\beta E)}
.
\end{align*}

  The process of collapse happens after even a single time-step
because the modeler's state of knowledge is now described by
a weighted set of possible wavefunctions, whereas the actual
ensuing dynamics will be associated with only one of those choices.

\begin{figure*}
\includegraphics[width=0.9\textwidth]{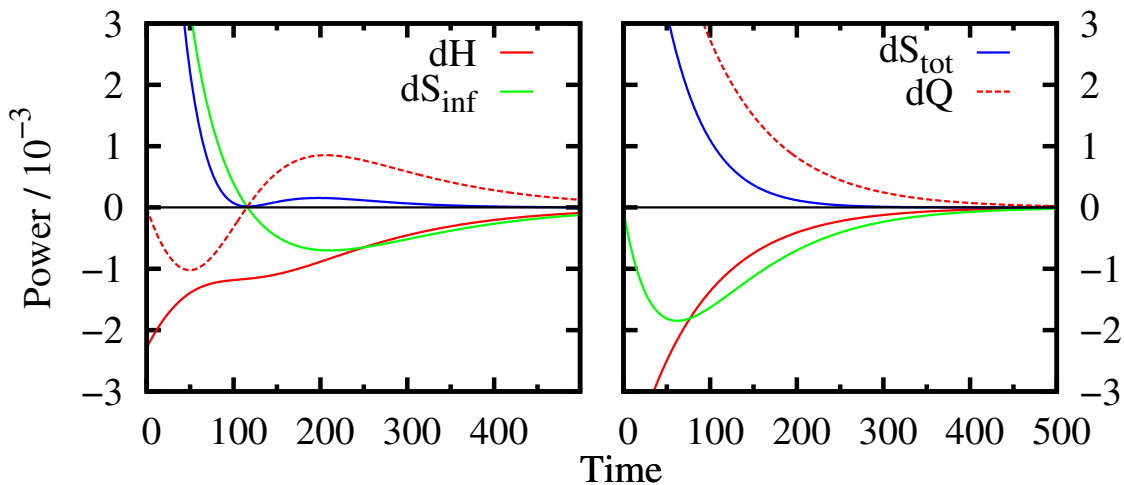}
\caption{ Entropy and heat production per unit time
from the spin system (Fig.~\ref{f:interfere}) during wavefunction collapse.
The left panel starts from the pure $x$-state, while the right panel
starts from the uniform directional distribution ($\rho = I_2$).
The `heat released' is defined as the difference, $dQ = dS_\text{tot} - dS_\text{inf}$.
Parameters: The oscillator energy was $\hbar\omega = 1$, with $\beta=2$.
The distribution for $r$ in $G^0$ was integrated over the full 3D normal distribution using
Gauss-Hermite quadrature and $D = 10^{-3}$.
}\label{f:ent}
\end{figure*}

  Figure~\ref{f:ent} shows the time-course of work and system entropy
for the dynamics of a spin system starting from the $\hat x$
superposition state (left panel), $\rho = \bigl[ \begin{smallmatrix} 1 & 1 \\ 1 & 1 \end{smallmatrix} \bigr]/2$,
and from a statistical mixture of up and down states (right panel),
$\rho = \bigl[ \begin{smallmatrix} 1 & 0 \\ 0 & 1 \end{smallmatrix} \bigr]/2$.
Energy and information entropy differences, $dH$ and $dS_\text{inf}$ at
each time step are differences between thermostatic state functions, $\Tr{H \rho(t)}$
and $\Tr{-\log \rho}$.  The total entropy increase ($dS_\text{tot}$, Eq.~\ref{e:S}) is
a measure of irreversibility, not a state function.  Due to its definition, it contains an
exact contribution, $dS_\text{inf}$.  Subtracting this gives a quantity which we term
$dQ$, to represent the heat given off to the environment.

  From the perspective of $z$ as the unique axis, both starting states look identical -- giving a 50\%
probability of detecting the starting state as `up.'  However, the information entropy
of the first matrix is the minimum possible (zero), while the second is the maximum possible
($\log 2$).  As time progresses in the left panel, the entropy of the state first increases
due to random collisions before finally decreasing to settle down into an equilibrium state.
However, the energy continually decreases as collisions tend to rotate the spin into the $z$ direction
on average.  This gives an initially negative `heat,' since knowledge of the pure starting state
could have been used to cool a properly constructed external environment.
after the cross-over region, the density matrix is compressed toward the up-state,
decreasing the information entropy and requiring a release of heat into the environment.

  The initial heating region is completely absent when starting from the uniform
distribution in the right panel of Fig.~\ref{f:ent}.  Since the system starts in the maximally
uncertain state, it continually gives off heat while lowering its energy.  Initially, the
compression (measured by decrease in information entropy) is slow with large
decreases in energy.  Eventually, the system settles down into a final compression
phase similar to the left panel.

  The dotted lines are plotted using the path relation for the total entropy derived in Eq.~\ref{e:S}.
It can be seen that the total entropy increase per unit time is always positive,
reflecting the trade-off between work and entropy.  Work has to be done
to compress the distribution toward the up-state.
In the absence of environmental noise, superposition makes it impossible
for this to happen.
Here, it occurs because our probability assignment for collisions accounts
for the fact that the energy of the system plus its environment
is conserved.  Because of this, the environment is not allowed to impart arbitrarily
large amounts of energy to the system.  The total entropy has to increase
when work is done in this `thermal' way, since the number of allowable
collision events decreases with time.

\section{ Discussion}

  It is at first surprising how closely the dynamics of the energy-weighted propagator
matches the model with an explicit cavity and measurement process.
Both show a nearly exponential decay toward an eventual steady state.
The steady-state is very nearly a Boltzmann distribution at high temperatures or weak coupling.
This result should be expected to hold independently from the specifics of the environmental
noise process.  The nonequilibrium dynamical method here thus fulfills its goal
of providing a first-principles model for the dynamic energy exchange processes
which {\em derives} the equilibrium distribution.  Both models also show deviations
toward higher effective temperatures due to environmental coupling -- an effect
observed in experiments.\cite{yutsu10}

  The low-temperature behavior of the stationary state is a physically
real effect.  It occurs here because of the continual random influence of the environment,
removing the possibility of approaching absolute zero temperature.
It will have an imperceptible effect on the energy-temperature equation of state, since
the behavior of the system is dominated by the ground state at these low temperatures.
Unless the interaction is unreasonably large, this method still gives the probability of being in
the ground state at 1 to $\sim$7 decimal places.

\subsection{ Relation to Entropy and Work Fluctuation Theorems}

  Both the definition (Eq.~\ref{e:S}) and its interpretation
differ from the usual fluctuation theorems.\cite{cjarz08,mcamp11}
Conventional fluctuation theorems are defined with respect to a
Boltzmann-like stationary distribution and an exact, reversible,
dynamical process (or a stochastic process
obeying detailed balance for that stationary distribution).
The transition entropy approach pursued here
begins by postulating a stochastic transition process and only
then determines its equilibrium and nonequilibriumn
properties.

  Because of this difference in origin, the entropy definition
also differs from other expressions related to the second law.\cite{gcroo99,cjarz08,sdeff10,tsaga13}
The origin of the usual fluctuation theorem in exact dynamics leads
to vanishing irreversible heat dissipation in the log-average,
\begin{align*}
dS_\text{tot} &= \avg{\log P(\Delta E)/\tilde P(-\Delta E)} \\
 &= \beta \avg{\Delta E - \Delta A} = \Delta S_\text{inf}
,
\end{align*}
where $A$ is the Helmholtz free energy of a thermostatic equilibrium state
at temperature $k_B/\beta$ allowed by the
dynamics.  Of course, the equilibrium relation between $A$ and $\avg{E} = U$
sets $dS_\text{tot} = \Delta S_\text{inf}$, and there is no irreversible entropy production.
In order to make statements about irreversibility, exact fluctuation relations have
to be supplemented by a model for the thermalization of the environment.  It
is the entropy increase of the environment that is ultimately responsible for
irreversibility derived in this way.

  Several work fluctuation theorems have been adapted to the quantum mechanical
setting using moment generating functions,
\begin{equation*}
\Tr{e^{-i\gamma H} U(t) e^{i\gamma H} \rho U^\dagger(t)},
\end{equation*}
for the energy difference between initial and final states.\cite{ptalk09,mcamp11}
These are closely related to the energy-weighted propagator derived here (Eq.~\ref{e:T2}).
They are the natural analogues to the definition of work distributions used for deriving the
classical fluctuation theorems.  The twist of perspective in this work shows that
their success can also be viewed as justification for postulating a  {\em statistical
law of random transitions}.  Viewed in this way, the exponential averages are not
just a calculational device for comparing to some
`instantaneous reference distribution' but are also useful statistical likelihoods
for observing transitions in energy space.  The results of this work show that
these likelihoods can be relied upon to generate a physical
steady-state distribution for an open quantum system -- without any
reference to the Boltzmann-Gibbs equilibrium.

\section{ Conclusions}

  This work provides a new perspective on the dynamics of open quantum systems.
Three complete examples were analyzed.
Understanding this perspective motivates a new statement of the
central question for nonequilibrium systems:
``What is the most logically consistent way to represent an experimenter's
subjective probability of the system-environment interaction when only
the system's trajectory is known?''
Utilizing a model of the maximum entropy type for this interaction gives
an implicit representation of the environment based only on its
role as a source of noise and dissipation.  In addition, the
dynamics is qualitatively very similar to results with a more explicit
representation of the environment.

  A second question was also considered, related to the most complete
description of the system in-between interactions with the environment.
The firmly established measurement hypothesis of quantum theory
was used to answer it.  A single wavefunction is a complete
description of an isolated system, but a density matrix represents
the maximal combination of physical and statistical information available for measurement.
Agreement with the peculiar property that individual wavefunctions
do not represent mutually exclusive events was used to select
an admissible form for the propagator.  The conclusion was to constrain
the energy change induced by the operator rather than the
energy change of the wavefunction itself.
Subtleties with normalization of nonlinear models based on
the density matrix lead to the final form for Eq.~\ref{e:T2}.

  A completely linear propagator, $G^{(2)}_N$ (Eq.~\ref{e:G2})
was mentioned as a possibility, but rejected because it
did not clearly connect initial and final wavefunction states
between time-steps.  Although its stationary state is simpler to
find numerically, preliminary results (not shown) display
a minimum effective temperature.  At very low temperature (high $\lambda$),
the stationary state probabilities start to distribute uniformly across
energy levels.

  The dynamics of the favored propagator, $G_N$ (Eq.~\ref{e:T2}), were compared to a reduced
model for thermalization of a trapped atom in an optical cavity.
In addition to showing similar relaxation rates and dynamical behavior,
the energy-weighted propagator showed mathematical properties
consistent with maximum entropy
thermodynamics.  Specifically, the transition free energy (Eq.~\ref{e:F})
acts as a generating functional for the distribution of energy flux (Fig.~\ref{f:work}).

  In addition, the transition entropy (Eq.~\ref{e:S}) describes the
trade-off between cooling of the system (lowering its information entropy)
and heat release into the environment.  Running a collapse experiment
on a spin in a pure, superposition state initially absorbs heat from the environment,
whereas the same experiment on a completely random spin continually
decreases entropy, requiring a continuous heat release into the environment (Fig.~\ref{f:ent}).
The physical manifestation of this difference is because incoming random light will be scattered
in a preferential set of directions by the known spin.
The transition entropy serves as a clue to this opportunity.
However, actually using this initial absorption of heat requires a
way to make use of the fact that the particle begins in a known state, e.g. by
capturing the scattered light.
Otherwise, the known force exerted by the particle is silently converted into heat.
Again, we find that entropy is a consequence of the interplay between the known
and unknown.

  This work presents a consistent picture of the role of the observer
and the environment throughout the process of wavefunction collapse.
Collapse is not possible without a noisy environment.  Interacting
with that environment, however, will cause further entanglement
rather than collapse.  Nevertheless, the interaction will remove 
coherences present in the system,
while favoring ending states that are lower in energy.
At this point, the experimenter's state of knowledge about the system
is described by a density matrix that accounts for these differences in energy.
However, the actual dynamics of the system will be due only to a single wavefunction
belonging to that density matrix.
The most logically consistent way to guess the system's state
after some amount of interaction has taken place is to
choose a wavefunction from the density matrix using the usual quantum measurement rule.
Put another way, wavefunction collapse never occurs, but the
experimenter's state of knowledge becomes heavily weighted toward
individual wavefunctions representing a collapsed state.
In this interpretation, Schr\"{o}dinger cat states are incomplete
until some environmental interaction takes place.

  The reduction of the environment to a statistical noise source with
a bare minimum number of properties is critical to achieving this picture.  Rather than starting
from an exact dynamical process for the environment, it is possible
to present a set of interaction Hamiltonians and derive a weighted propagator.
It is immediately clear how to add
time- and spatially-dependent models for the environment (e.g. correlated noise sources).
Other maximum entropy-type constraints, e.g. the change in the total momentum, $\sum_i \hat p_i$,
can also be included in the biasing exponential along with the energy weight.
The result is a greatly simplified model of the essential elements for nonequilibrium
quantum systems.


\section*{Appendix}

\section{Comparison with the Density Matrix Master Equation}

  This section shows that another well-known formalism for simulating
the dynamics of open quantum systems can be derived as a low-order
approximation of the energy-weighted propagator (Eq.~\ref{e:T2}).

  Master equations for the density matrix are the most popular means for
describing the role of environmental noise in localizing particles on a microscopic
scale.\cite{acald81}  Caldeira and Leggett began with a
description of collapse of a density toward the diagonal induced by coupling
to a bath of harmonic oscillators.  However, this coupling alone
does not provide a suitable mechanism for energy dissipation.  Later, these
decoherence models were extended to add dissipation
by imposing an energy distribution on the bath variables.\cite{acald83,wzure86}

  The particular high-temperature limit derived by Caldeira and Leggett
for external forces linear in $x$ is,\cite{ejoos98,wzure02}
\begin{align}
\dot \rho &= -\frac{i}{\hbar} [H, \rho] + \frac{i\gamma}{\hbar} [\{\hat p, \rho\}, x] + \frac{2m\gamma}{\beta \hbar^2} (x \rho x - \frac{1}{2} \{x^2, \rho\}) \label{e:master} 
 ,
\end{align}
where the damping rate should be of dimension $\gamma \sim \sqrt{c/2m}$.

  It will be shown that the terms of Eq.~\ref{e:master} are in 1:1 correspondence
with those appearing in the high-temperature limit of $G^{(2)}_N [\rho]$.
In this limit, $G_N$ and $G^{(2)}_N$ are identical.

  The first step is approximating the energy weighting on $U_r$.
Since the random forces, $r$, are of order $\sqrt{\Delta t}$, terms up to $r^2$ contribute,
\begin{align}
e^{-\lambda\hat H/2} e^{-\frac{i\Delta t}{\hbar} (\hat H - r x / \Delta t)} e^{\lambda\hat H/2}
\equiv W(e^{-\frac{i\Delta t}{\hbar} (\hat H - r x / \Delta t)}) \\
\simeq W(I - \frac{i\Delta t}{\hbar} \hat H + \frac{irx}{\hbar} - \frac{r^2 x^2}{2\hbar^2})
.
\end{align}
When applying $G[\rho]$, the sign of $\lambda$ is switched on the right of
$\rho$.  This is denoted below by $\bar W$, so
\begin{equation}
G^{(2)}_N[\rho] = \int P^0(r) W(U_r) Z^{-1/2} \rho Z^{-1/2} \bar W(U_r) \; dr
.
\end{equation}

  The $r^\text{th}$ contribution to $G[\rho]$ becomes, for small $\Delta t$
and dropping terms linear in $r$ (which vanishes on averaging),
\begin{align}
 &W(I - \frac{i\Delta t}{\hbar} \hat H + \frac{irx}{\hbar} - \frac{r^2 x^2}{2\hbar^2}) Z^{-1/2} \notag \\
  & \qquad \rho Z^{-1/2} \bar W(I + \frac{i\Delta t}{\hbar} \hat H - \frac{irx}{\hbar} - \frac{r^2 x^2}{2\hbar^2}) \\
&\simeq \rho - \frac{i \Delta t}{\hbar}[\hat H, \rho] + \frac{r^2}{\hbar^2} W(x)\rho\bar W(x) \notag \\
 &\qquad - \frac{r^2}{2\hbar^2} \left(W(x^2)\rho + \rho \bar W(x^2)\right) - \frac{\Delta t}{2} \left\{ \pd{Z}{\Delta t}, \rho \right\}
\end{align}
Of course, $W(I) = I$ and $W(\hat H) = \hat H$.

  Taking the high-temperature limit allows a first-order expansion of the
damped position operator,
\begin{equation}
W(x) \simeq x - \frac{\lambda}{2} [\hat H, x] = x + \frac{i\lambda\hbar}{2 m} \hat p
\end{equation}
Similarly, $W(x^2) \simeq x^2 + \frac{i\lambda \hbar}{2m}\{x, \hat p\} $.

  The expansion of $Z = \int \bar W(U_r^\dagger) W(U_r) P_0(r) dr$ is
just as for $G[\rho]$ above.  Defining the variance as
$\int r^2 P_0(r) dr = \sigma^2 \Delta t$,
\begin{align}
Z &\simeq I + \frac{\sigma^2 \Delta t}{\hbar^2} \bar W(x) W(x) - \frac{\sigma^2 \Delta t}{2\hbar^2}(W(x^2) + \bar W(x^2)) \notag \\
 &= I + \frac{\sigma^2 \Delta t}{\hbar^2} \frac{i\lambda\ \hbar}{2 m} (x \hat p - \hat p x) = I - \frac{\lambda \sigma^2\Delta t}{2 m}
\end{align}

  Completing the calculation above gives
\begin{align}
\dot \rho = - \frac{i}{\hbar}[\hat H, \rho]
&+ \frac{\sigma^2}{\hbar^2} \Big( x\rho x - \tfrac{1}{2}\{x^2, \rho\} \\
 &                     + \frac{i\lambda\hbar}{2m} \big(\hat p \rho x - x\rho \hat p
                      					- \tfrac{1}{2}[\{x, \hat p\}, \rho]\big)
                                 \Big)
+ \frac{\lambda\sigma^2}{4m} \rho \notag
.
\end{align}
Rearranging $[\{x, \hat p\}, \rho] = 2(x\hat p \rho - \rho\hat p x - i\hbar\rho)$ cancels the constant.  The expression multiplying $i\lambda$, responsible for dissipation, becomes $[\{\hat p, \rho\}, x]$,
yielding a final expression identical to Eq.~\ref{e:master}.  The identifications can be made,
\begin{align}
\sigma^2 &= 2m\gamma / \beta \\
\gamma &= \lambda\sigma^2/2m
\end{align}
This model therefore corresponds to a high-temperature limit
of Eq.~\ref{e:maxent2}, where $\lambda = \beta$.
Numerical results in this work confirm that the steady state for Eq.~\ref{e:maxent2}
at low temperature is $\lambda = \beta/2$.  The temperature discrepancy
appears to be due to dropping higher-order terms in $\beta$.

\end{document}